\documentclass[%
 aip,
 amsmath,amssymb,
preprint,%
nofootinbib
]{revtex4-1}

\usepackage{amsmath}    
\usepackage{graphicx}   
\usepackage{verbatim}   
\usepackage{color}      
\usepackage{subfigure}  
\usepackage{hyperref}   
\usepackage{soul}
\usepackage{hyperref}
\usepackage{tabularx}

\usepackage{float}
\usepackage{amssymb}
\usepackage{mathtools}
\usepackage{setspace}
\usepackage{amsmath}
\usepackage{amsfonts}
\usepackage{bm}
\usepackage{outlines}
\usepackage{xcolor,cancel}

\usepackage[utf8]{inputenc}
\usepackage[T1]{fontenc}
\usepackage{mathptmx}

\usepackage{dcolumn}
\usepackage{float}
\usepackage{ulem}

\usepackage{tcolorbox}
\usepackage{graphicx}
 \usepackage{multirow}
\usepackage{hyperref}
\hypersetup{colorlinks,allcolors=blue}
\usepackage{blindtext}
\definecolor{color}{RGB}{0,0,0}
\begin{document}

\title{Recharging and rejuvenation of decontaminated  N95 masks}
\date{\today}

\author{Emroj Hossain}
\affiliation {Department of Condensed Matter Physics and Materials Science, Tata Institute of Fundamental Research, Mumbai 400005, India}

\author{Satyanu Bhadra}
\affiliation {Department of Condensed Matter Physics and Materials Science, Tata Institute of Fundamental Research, Mumbai 400005, India}

\author{Harsh  Jain}
\affiliation {Department of Condensed Matter Physics and Materials Science, Tata Institute of Fundamental Research, Mumbai 400005, India}

\author{Soumen Das}
\affiliation {Department of Condensed Matter Physics and Materials Science, Tata Institute of Fundamental Research, Mumbai 400005, India}

\author{Arnab Bhattacharya}
\email{arnab@tifr.res.in}
\affiliation {Department of Condensed Matter Physics and Materials Science, Tata Institute of Fundamental Research, Mumbai 400005, India}

\author{Shankar Ghosh }
\email{sghosh@tifr.res.in}
\affiliation {Department of Condensed Matter Physics and Materials Science, Tata Institute of Fundamental Research, Mumbai 400005, India}

\author{Dov Levine}
\email{levine@technion.ac.il}
\affiliation { Department of Physics, Technion-IIT, \textcolor{color}{32000} Haifa, Israel}

\date{\today}

\begin{abstract}
N95 respirators comprise a critical part of the personal protective equipment used by frontline health-care workers, and are typically meant for one-time usage. However, the recent COVID-19 pandemic has resulted in a serious shortage of these masks leading to a worldwide effort to develop decontamination and re-use procedures. A major factor contributing to the filtration efficiency of  N95 masks is the presence of an intermediate layer of charged polypropylene electret fibers that trap particles through electrostatic or electrophoretic effects. This charge can degrade  when the mask is used.  Moreover, simple decontamination procedures (e.g. use of alcohol) can degrade any remaining charge from the polypropylene, thus severely impacting the filtration efficiency post decontamination.
In this  report, we summarize our results on the development of a simple laboratory setup allowing measurement of charge and filtration efficiency in N95 masks. In particular, we propose and show that it is possible to recharge the masks post-decontamination and recover filtration efficiency.
\end{abstract}

\maketitle

\textcolor{color}{Face masks  are  our first line of defence against  airborne particulate matter \cite{dbouk2020respiratory, verma2020visualizing}}. In particular N95 \footnote{Athough we shall use the term N95, our results pertain equally to PPF2 and KN95 respirators which work on the same basis.  Moreover, we note that the protocol described herein works for surgical masks and other filtering facepiece respirators as well. } respirators comprise a critical part of the personal protective equipment (PPE) used by frontline health-care workers \textcolor{color}{as they  provide a barrier for transmission of pathogen laden droplets that are ejected  by  coughing, sneezing, talking or breathing  by an infected person \cite{busco2020sneezing,brosseau2009n95,centers2020decontamination,dbouk2020coughing}}. The name designation N95 indicates that these masks can filter $ 0.3 \mu$m sized particles with $ 95\% $ efficiency \cite{brosseau2009n95}.  N95 masks are meant for one-time usage for two reasons: (1) potential contamination and (2) rapid degradation of their filtration efficiency with use. However, the recent COVID-19 pandemic has resulted in a serious shortage of these masks which has started an intensive search for methods which would allow for multiple use.

Most of the literature has dealt with various
proposals for decontamination procedures, including careful use of dry and wet heat or exposure to hydrogen peroxide vapor, ozone, UV radiation, or alcohol \cite{centers2020decontamination,fischer2020assessment,liao2020can,kumar2020n95, 2020decontaminating,li2020s, mackenzie2020reuse, ma2020decontamination, o2020efficacy}. 
While each of these methods likely deactivates viruses, it seems to be common knowledge that such  procedures adversely impact filtration efficiency and may even cause deterioration of the structural integrity of the mask.  

Less attention has been focused on restoring the filtration efficiency of a mask once it has become degraded; this is the question we address in this work.  In this paper we propose a method which, provided the mask has not been structurally compromised, can restore filtration efficiency to out-of-box levels.

\begin{figure}[t]
	\centering
	\includegraphics[width=1\linewidth]{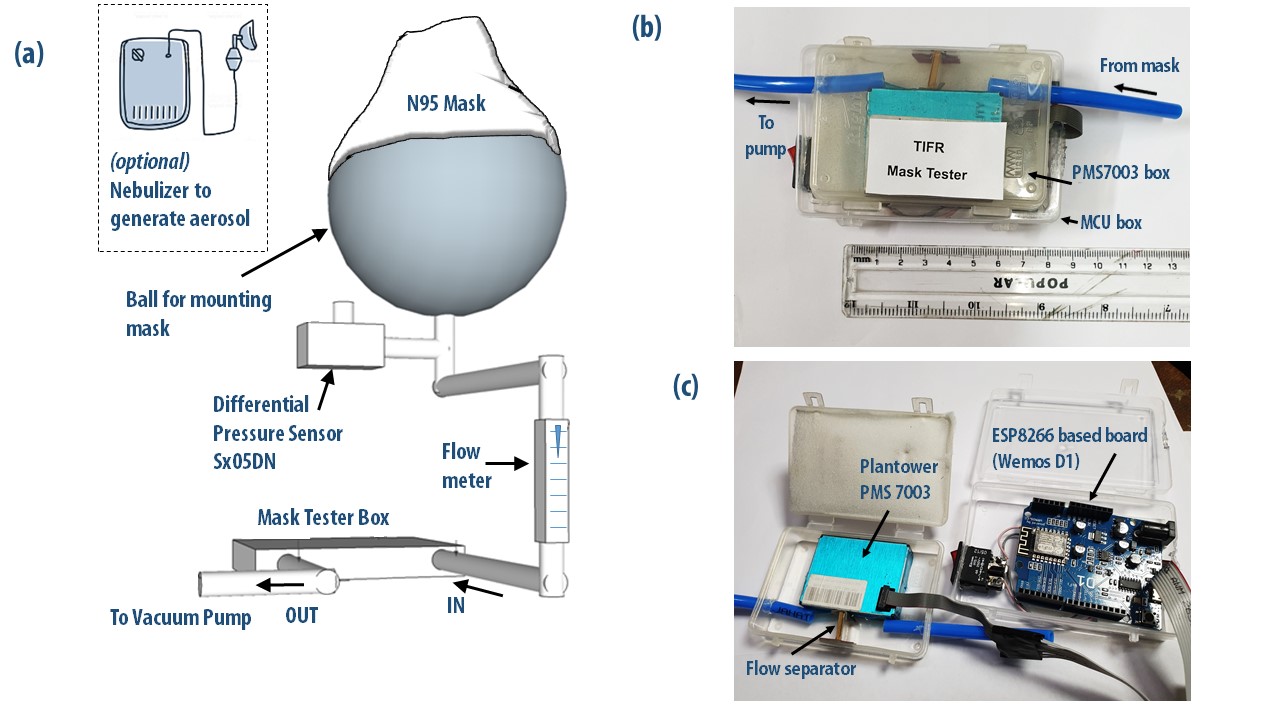}
	\caption{ \textcolor{color}{(a) Schematic diagram of the compact, low-cost mask tester developed in the lab. The mask is attached to a hard plastic ball simulating a human head, and air flow through the mask was effected by a small diaphragm pump. Particle counts were performed  by the using a Plantower PMS7003 air quality sensor chip  which  was interfaced to a  ESP8266 WiFi micro controller unit. (b) View of the air quality sensor chip and control unit in small plastic boxes kept atop each other in a compact configuration. (c) Opened up view showing the PMS7003 chip and the ESP8266 based MCU board. The box edges and connector ports are  hermetically sealed  to ensure that the setup is airtight. 	Additional  details of the experimental setup can be  found  in the \texttt{GitHub}   repository \cite{MAsk_tester}.}} 
	\label{fig:masktester}
\end{figure}

As with other filtration processes, N95 masks intercept foreign particles in different layers of the mask material. A particle can be captured either {\it mechanically} - if it encounters a mask fiber directly in its path - or if the mask material is such that it can attract and ensnare particles, say {\it electrostatically} \cite{thakur2013electret}. \textcolor{color}{On making contact with the surface of the  fibre, adhesive forces, such as the Van der Waals force, immobilize the particle on the surface of the fiber\cite{kumar2013weak}.}

Flow through a mask is usually thought to be laminar, such that the flow would usually bend smoothly around an obstacle (fiber). If this is the case, mechanical capture of the particle on the surface of the fiber happens when a particle deviates from its streamline path, causing an impact with the mask material.  This can happen for larger particles whose inertia is large enough to cause such a deviation from the streamline, or for smaller particles whose Brownian diffusion is strong enough \cite{finlay2001mechanics}.  \textcolor{color}{For filters based on fibrous materials, and operating at filtration velocities similar to those encountered in human breathing, the minimum filtration efficiency occurs for $\approx 0.3 \mu m$  sized particles. At this scale the filtration mechanism crosses over from a diffusion dominated regime  to   an inertia dominated regime \cite{lee1980minimum}}.

In addition to mechanical capture, N95 respirators employ an electrostatic mechanism to attract and intercept foreign particles (charged or uncharged). This happens when there are significant electric fields and electric field gradients in the mask material, which may occur when the fibers are charged \cite{frederick1974some}. It is these electrostatic interactions which raise the filtration of N95 masks to the 95\% level. Charged fibers can attract both inherently charged particles by Coulombic forces as well as neutral polar particles (such as tiny aqueous droplets) by dielectrophoretic forces that come from the interaction of polarized objects and electric field gradients.

In typical N95 masks, the electrostatic filtration is performed by a layer comprised of a  non-woven melt-blown mesh of  charged polypropylene  fibers. Most of the pores in this mesh have a characteristic length scale of about $ 15 \mu m $ and about $ 90\% $ of its space is void. This layer is held in place between two or more quasi-rigid layers that provide both support and mechanical filtration. \textcolor{color}{Polypropylene is an {\it electret}, a dielectric material which can hold a charge or possess a net microscopic dipole moment \cite{sessler1980physical}.}

Pure polypropylene is a non-polar polymer with a band gap of 8eV. However, the presence of molecular level defects both chemical and physical in nature allow the formation of localized energy states that can trap charge \cite{sessler1980physical}. Moreover, its electrical polarization properties are often enhanced by introducing various  additives like magnesium stearate \cite{zhang2018design} or $\text {BaTiO}_{3}$ \cite{kilic2015electrostatic}  which are added to the polymer melt to increase the electret performance. Even then, the charge on the polypropylene fibers undergoes significant degradation when open to the surroundings, which is exacerbated by the warm humid environment created by respiration during use. Additionally, most decontamination methods remove all the charges from the charged layer,  with a concomitant reduction in mask efficiency.

Thus, a key aspect of the performance of an electret-based mask is its ability to maintain its charge in a hot and humid atmosphere.  Failing this, extended usage can only be obtained through a cycle of decontamination and recharging, if this is possible.  It follows that a simple procedure for electrically recharging a decontaminated mask without disassembling it would be very useful, especially if it does not rely on special-purpose equipment which would not be readily available.

The standard methods for charging polymer fibers are
corona discharge \cite{pai1993physics}, photo-ionisation induced by particle beams (gamma rays, x-rays, electron beams) \cite{gross1962gamma, gross1958irradiation}, tribo-electrification \cite{mccarty2007ionic,zhao2020household,konda2020aerosol}, and
liquid contact charging \cite{chudleigh1973stability}.
These  methods are not easily deployable in hospital conditions on preassembled masks. In this note, we propose a simple recharging method based on high electric fields, and demonstrate its effectiveness. 

 \textit{Crucially, our method can be performed using readily available equipment and materials, and so can be employed both in urban and rural settings.}

\subsection*{Mask filtration testing setup}


Because of the COVID-19 pandemic, we did not have access to special-purpose mask filtration equipment, so we designed and constructed a rough apparatus to measure the efficiency of filtration of particulate matter, using an air-quality monitor as a particle counter.  The setup is shown in Figure \ref{fig:masktester}.   A plastic ball serves as our proxy of the human face, on which we place the mask that we want to test. Air is sucked through the mask  with a vacuum pump whose flow rate is controlled and monitored by a  flow meter.  \textcolor{color}{We use an oil-free diaphragm pump (HSV-1, High Speed Appliances, Mumbai) that provides a maximum flow of 30 lpm.
The flow can be measured and controlled with a taper-tube flow meter. For most experiments we used a flow of 10 lpm, similar to typical human breathing rates. }
This air is made to flow through an  particle counting setup  which contains a Plantower PMS7003 sensor \footnote{Typically,  particle counters  work  by analysing the light scattered by the particles.   Since the  size of the particles are comparable to the wavelength of  visible light, this scattering is of {Mie} type. Further, it is assumed that these are single scattering events.}.  \textcolor{color}{The details of the experimental setup can be  found  in the \texttt{GitHub}  repository \cite{MAsk_tester} ( see supplementary section}.

While the particle sensor chips are optimized for 2.5 $ \mu $m particle measurements, the Plantower PMS7003 sensor also has a 0.3$ \mu $m channel. The filtration efficiency ($\eta$) is determined from the ratio of 0.3$ \mu $m particles per unit time detected with the mask attached ($ N_{mask}$) to that without the mask attached ($ N_{ambient} $) as
$$
\eta =  \left(1-\frac{N_{mask}}{N_{ambient}}\right)\times 100.
$$
Our measurements are taken using a small diaphragm pump to suck air through a mask attached to a plastic ball at flow rates ($\sim 10 \, \mathrm{lpm}$)  of the order of physiological breathing rates. \textcolor{color}{Much higher flow rates ($\sim 80$ lpm) are often used to certify N95 masks.  To check for the  dependence of the filtration efficiency on flow rates, we measured filtration efficiency for flow rates between 3 and $30 \,\, \mathrm{lpm}$, and have found that the difference in the measured $\eta$ is about one percent}.  The efficiency of our particle counter for smaller particles is of order 50\%. Since $\eta$  is  related to the ratio of $N_{mask}$ and $N_{ambient}$, it is insensitive to the fact that not all the particles at $0.3 \mu m$ are  being counted. \textcolor{color}{We have cross-checked  the measurements  obtained with the Plantower chip with  a Lighthouse  clean-room particle counter, and we found the measurements of $\eta$ by both the devices to be consistent.} For the ambient air to be filtered, we generated aerosols of normal saline solution (0.9\%) by employing a standard medical  nebulizer. \textcolor{color}{ These nebulizers produce a  broad distribution of droplet sizes ranging from 100 nm to 10 $\mu $m \cite{ferron1997estimation}.  }
 
\begin{figure}[t]
	\centering
	\includegraphics[width=.9\linewidth]{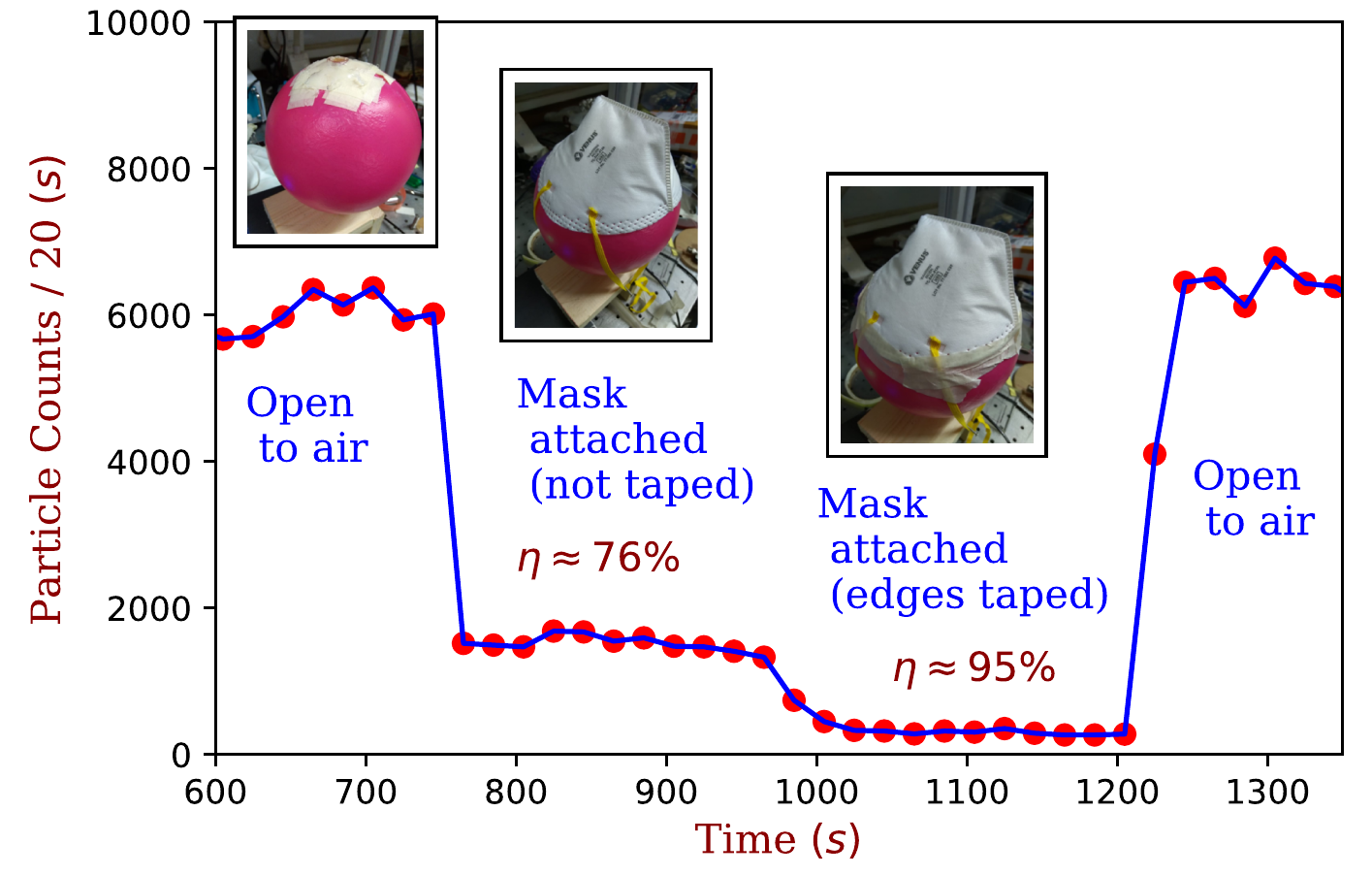}
	\caption{Filtration tests on a pristine  Venus 4400 N95 mask. For the initial and final readings, no mask was attached, and so serve as a baseline. When untaped, the seal between the mask and the ball is imperfect, and the filtration efficiency is $76\pm 1\%$. Upon taping the mask to the ball, we obtain $\sim 95\pm 1\%$ filtration efficiency.
	}
	\label{fig:masktester_b}
\end{figure}

 The fit of the mask to the plastic ball is imperfect, allowing air leakage from the sides. To obtain reproducible values, the masks edges were taped to the ball using paper masking tape. The filtration data, albeit employing a home-made testing apparatus, should be sufficient to make at least semi-quantitative comparisons between one mask and another, and quantitative comparisons between the same mask in its charged and uncharged states. To give a sense of the measurement, the  filtration data from a pristine N95 mask is shown in Figure \ref{fig:masktester_b}. When the mask is placed on the ball without taping the sides, its efficiency was $76 \pm 1\% $.  Upon taping the sides, the efficiency improved to  $95\pm 1\% $. \textcolor{color}{ The reduction in filtering efficiency due to poor fitting is  a generic problem associated with the use of face masks \cite{dbouk2020respiratory}.}

\subsection*{Charge measurement}

\begin{figure}[t]
	\centering
	\includegraphics[width=.85\linewidth]{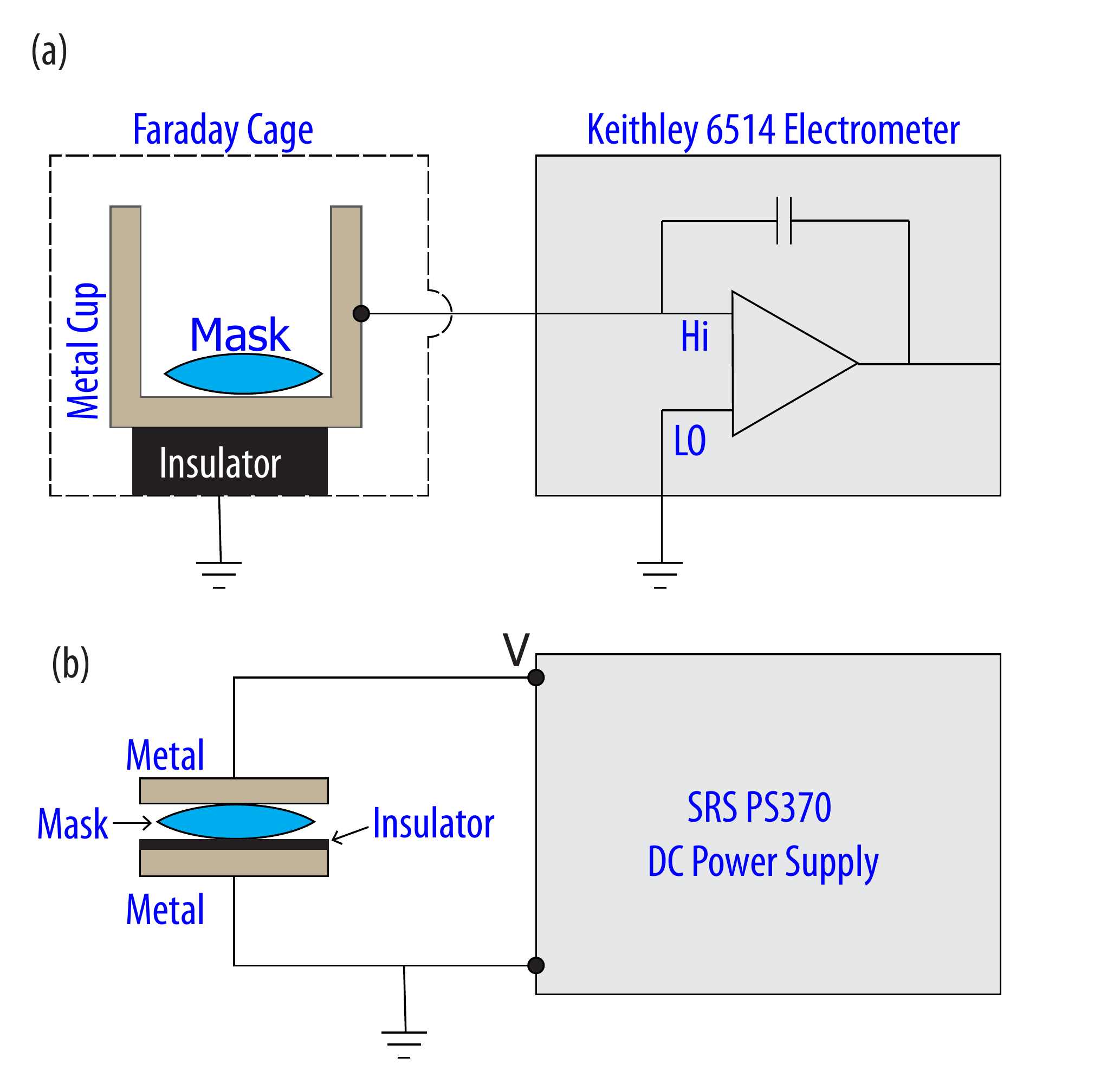}
	\caption{Schematic of charge measurement setup (a)  and mask recharging (b). In the mask recharging setup  a $ 100 \mu $m thick insulator (PET plastic sheet) was inserted between the mask and the ground electrode.  \textcolor{color}{Under high field the mask acts like an electrode on which the charges can be deposited.  The insulator allows the mask to get charged because it prevents any current from flowing in the circuit.}  }
	\label{fig:exptsetup}
\end{figure}

As shown in Figure \ref{fig:exptsetup}(a), We used a Keithley 6514 electrometer to measure the charge, with the mask placed in a metal cup which was electrically isolated from the ground by  an insulating teflon surface. The input of the electrometer is a three-lug triax connector, with the innermost wire (input high) being the charge sensing terminal. This charge sensing terminal of the electrometer was connected to the  metal cup.   In our experiments we used the guard-off condition, i.e., the common (input low) and the chassis are grounded.  Electrometers measure charge  by transferring the charge from the point of measurement to the reference capacitor of the electrometer, and only free charge can be transferred. Therefore, since it does not account for any bound charge, our measurement likely underestimates the total charge on a mask and so should be regarded as giving a relative indication rather than a precise measurement of the total charge on the masks.
This being said, there appears to be a  \textcolor{color}{\textit{qualitative}} correlation between measured charge and filtration efficiency, with masks with higher values of measured charge having higher filtration efficiency $ \eta $ (see Table \ref{tab:charge}). \textcolor{color}{ The data  both N95 and surgical masks are tabulated in the table. The surgical masks are different than the N95 masks in construction. Hence,   comparisons between charge and filtration efficiency should be made between masks of the same type. Moreover, we our charge measurement  technique is not sensitive to the dipolar character of the electrets.  Hence  quantitative calculation of correlation based on free charges  can not be estimated from this measurement alone.}

\begin{table}[t]
\begin{tabular}{|l|l|l|l|}
\hline
Brand Name                                                             & MaskType & \begin{tabular}[c]{@{}l@{}}Filtration\\ Efficiency\end{tabular} & \begin{tabular}[c]{@{}l@{}}Charge\\ (nC) \end{tabular} \\ \hline
O\&M Halyard 4627     & N95      & $98 \pm 1\%$   & $9\pm 0.5$  \\ \hline
Venus-V4420           & N95      & $96 \pm 1\%$   & $1 \pm 0.5$  \\ \hline
Primeware Magnum      & N95      & $95 \pm 1 \%$   & $1 \pm 0.5 $  \\ \hline
K95                   & N95      & $95 \pm 1 \%$   & $1 \pm 0.5 $  \\ \hline
\begin{tabular}[c]{@{}l@{}}
Magnum Viroguard \\ 
FFP1 3-ply
\end{tabular} 
                     & Surgical  & $98 \pm$\%   &$8\pm 0.5$  \\ \hline
Magnum SMS 3-ply     & Surgical  & $79 \pm 1\%$ &$2.9\pm 0.5$ \\ \hline
Magnum 3ply          & Surgical & $65 \pm 1\% $ & $1.3 \pm 0.5 $\\ \hline
\end{tabular}
\caption{Filtration efficiency and charge of masks tested. The surgical masks are different than the N95 masks in construction, thus, the comparisons between charge and filtration efficiency should be made between masks of the same type. \textcolor{color}{The error in the  charge measurement is mainly statistical in nature and comes from the uncertainty in the contact between the mask and the electrode. Note that the resolution of the charge measurement capability of the Keithley 6514 electrometer is few femto coulombs, which is orders of magnitude smaller than what is measured}. }
\label{tab:charge}
\end{table}

\subsection*{Recharging}

The masks were recharged by sandwiching them between two metal plate electrodes, which were connected to the high and the low output terminals of a SRS PS370 power supply. The low output terminal was grounded and a suitable voltage of positive or negative polarity was applied from the high output terminal of the source meter; Figure \ref{fig:exptsetup}(b) sketches the recharging setup. 

Our recharging method exploits the nonlinear conductivity of electrets, in particular polypropylene, as a function of applied electric field.  The electrical conductivity of polypropylene is dominated by hopping  \cite{foss1963electrical,okoniewski1994hopping}. Thus  at high fields, the conductivity of polypropylene is high, which makes the introduction of excess charges into the material possible by connecting it to a charge source.  

\begin{figure}[t]
	\centering
	\includegraphics[width=1\linewidth]{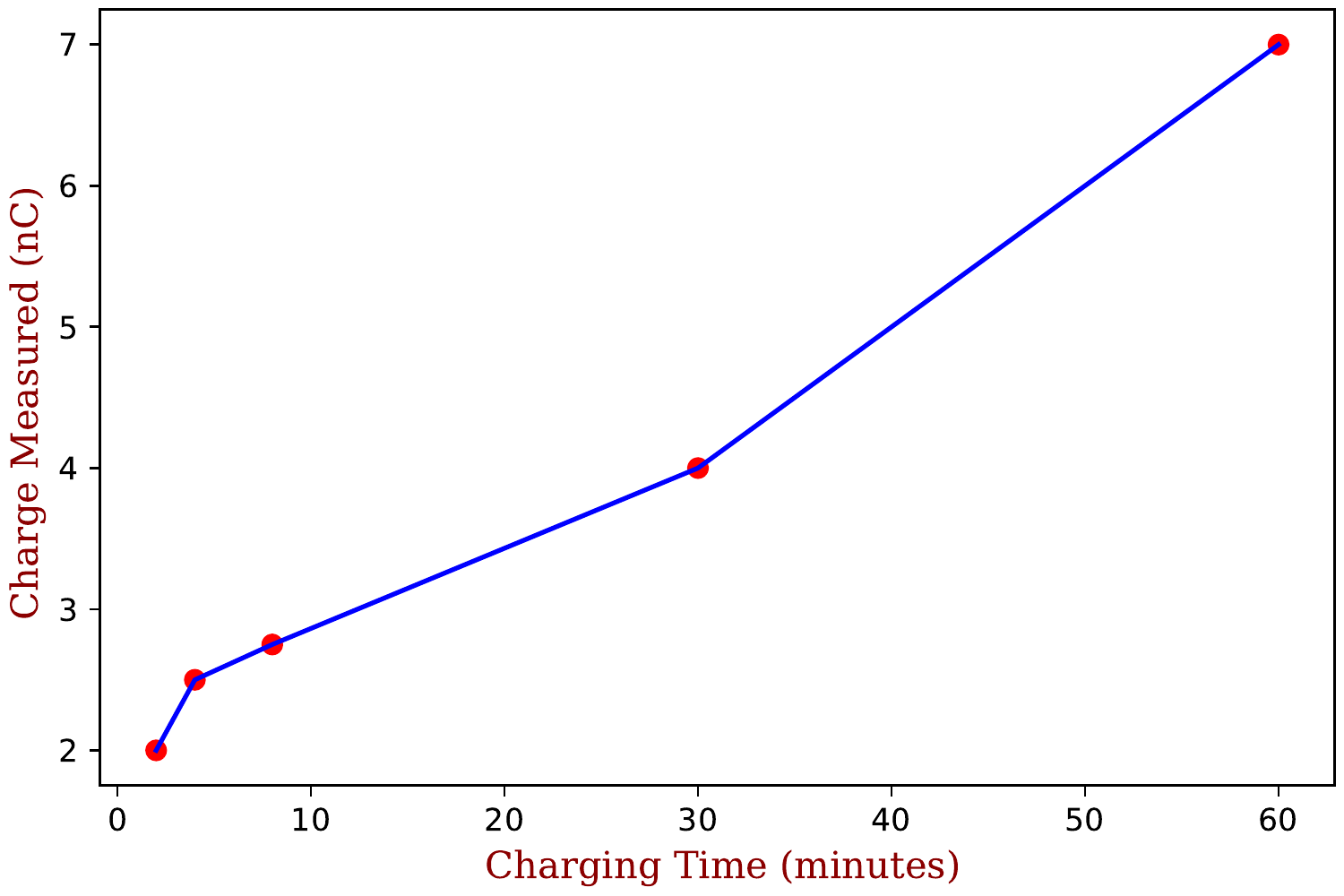}
	\caption{Charge accumulated on an O\&M Halyard 4627 mask as a function of charging time at 1000V. The charge on the pristine mask was $\sim 8$ nC. }
	\label{fig:chargingparam}
\end{figure}

When the charge source is switched off, the applied electric field becomes zero, and conductivity of the polypropylene drops effectively to zero. As a result, the added charge carriers become immobile, and the material remains charged. We find that the total charge deposited on the masks depends strongly on the charging time, as seen in Figure \ref{fig:chargingparam}, which shows the result of different charging times on a N95 mask, with the pristine value almost reattained after a 60 minute charge at 1000V. 

\subsection*{Filtration efficiency of recharged masks}

In the previous section we demonstrated that the application of relatively high voltage recharges the masks.  Of course, the important test is whether this recharging translates into improved efficiency in the filtration of fine particles.  To assess this, we first obtained a baseline measurement for the filtration efficiency of new unused masks\footnote{\textcolor{color}{The masks were not individually vacuum sealed. }}.  We then performed typical sanitization protocols, during which the masks typically lost most of  their charge\footnote{\textcolor{color}{We emphasize again that we are measuring primarily the free charge.}}, and measured the filtration efficiency of the discharged masks.  We then recharged the masks and measured their filtration efficiency. The  effect of  different  sanitization protocols and  recharging on the filtration efficiency of the masks is tabulated  in Table \ref{tab:caption}.

\begin{figure}[b]
	\centering
	\includegraphics[width=1\linewidth]{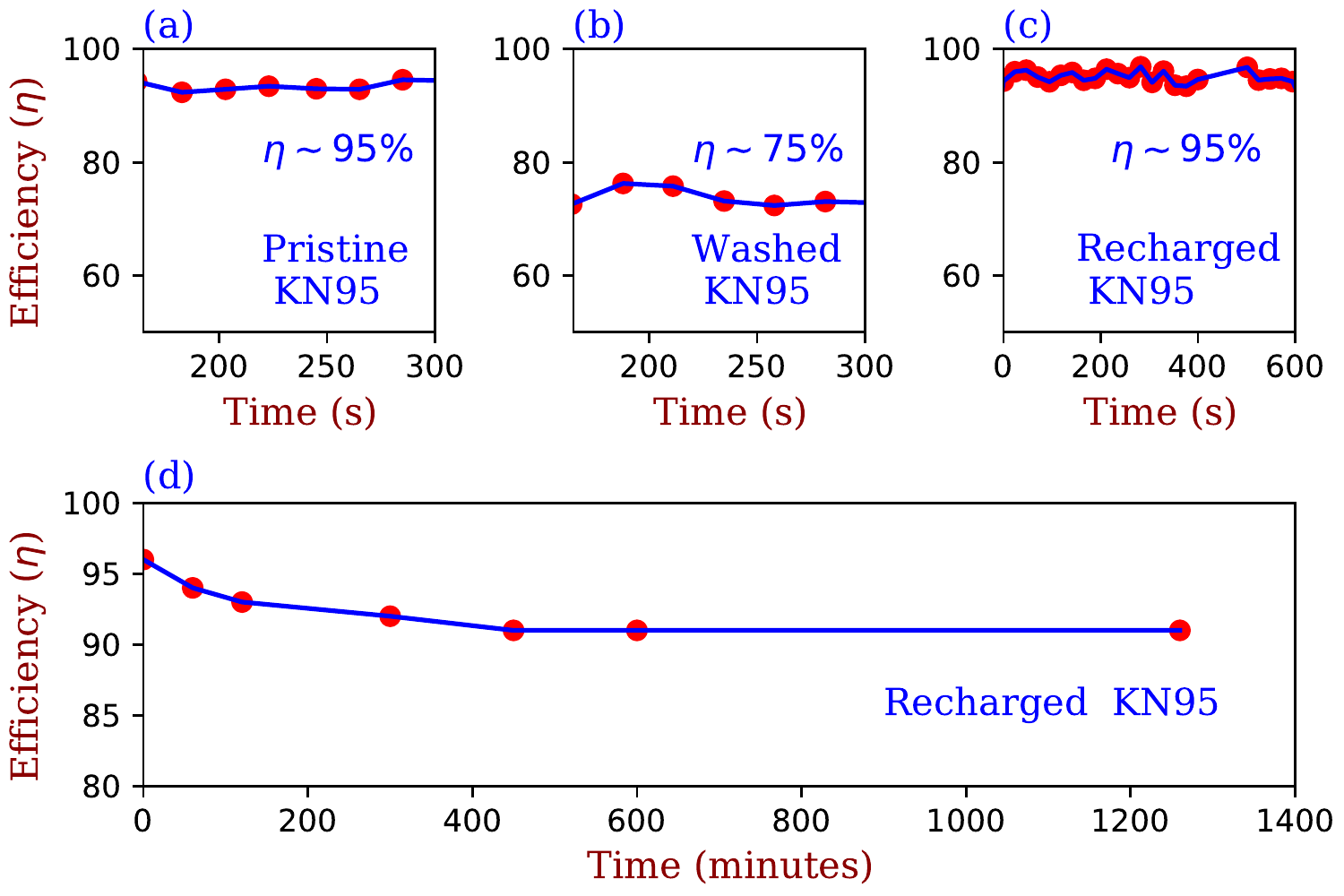}
	\caption{Top panel: (a) Comparison of the filtration efficiency of a new KN95 mask ($ 95\pm 1\%$),  (b) the same mask after washing and drying ($75\pm 1 \%$), and (c) the same mask after recharging for 60 minutes ($95 \pm 1 \%$). The filtrations  measured are for 0.3$ \mu m $ sized particles.  The  top rightmost panel of the figure shows that the filtering efficiency  $\eta$  is unchanged over a time span of   10 minutes.  The bottom panel (d) shows the decay of the efficiency of the recharged mask over the course of a day. } 
	\label{fig:rechargen95recharge}
\end{figure}

\begin{table}[t]
	\begin{tabular}{|l|l|l|l}
		\hline
		\multirow{3}{*}{\begin{tabular}[c]{@{}l@{}} Sanitization
		\\ Method \\ (Mask Brand) \\ \end{tabular}} & \multicolumn{2}{l|}{Filtration Efficiency ($\eta$) }                                                                                       \\ \cline{2-3} 
		& \begin{tabular}[c]{@{}l@{}}Before\\ Recharge\end{tabular} & \begin{tabular}[c]{@{}l@{}}After\\  Recharge\end{tabular} \\ \hline

		\begin{tabular}[c]{@{}l@{}} Ethanol\\ KN95\end{tabular}
		                                                                    &    $90\pm 1\%$                                                       & $96 \pm 1\%$                                                       \\ \hline

		\begin{tabular}[c]{@{}l@{}} Boiling Water\\ KN95\end{tabular}                                                                   & $74 \pm 1\%$                                                       & $86\pm 1\%$                                                        \\ \hline
		\begin{tabular}[c]{@{}l@{}} Washing Machine\\ KN95\end{tabular} 
		                                                                 & $75 \pm 1 \%$                                                      & $ 95 \pm 1\%$                                                     \\ \hline
		
		\begin{tabular}[c]{@{}l@{}}  Steam Exposed\\ (Venus V-4420N95)  \end{tabular}                                                                   & $77 \pm 1 \%$                                                       & $86  \pm 1 \%$                                                  \\ \hline
		\begin{tabular}[c]{@{}l@{}}  Ethanol\\ (Magnum N95)  \end{tabular}                                                                  & $ 50 \pm  1\%$                                                     & $ 86 \pm 1 \%$                                                      \\ \hline
		
	\end{tabular}
\caption{ 
The drop in the filtration efficiency in N95 masks  due to different protocols of decontamination and the  recovery of filtration  by recharging the masks. For the KN95 masks we used 2000V while for the Venus and Magnum N95masks we used 1000V \footnote{We observed an increase in the efficiency by few percentages on increasing the recharging voltage from 1000V to 2000V. However, beyond 2000V  further increase in the applied voltage did not alter the efficiency of the mask appreciably.}.  For decontamination by ethanol,  a new KN95/Magnum mask was soaked  in ethanol and then dried.  In the boiling method, the  mask was  immersed in boiling water for 1 hour and then dried. For the washing machine method the mask was laundered in a regular washing machine in a standard 40 $^{\circ}$ C, 84 minute cycle, wash/rinse/spin dry cycle.   For the steam method, the Venus mask was exposed to steam for 5 minutes on each side. For all these protocols we started with a new N95mask.
}
\label{tab:caption}
\end{table}

Representative data of the filtration efficiency of various masks after decontamination and recharging is given in the top panels of  Figure \ref{fig:rechargen95recharge}, where we start with a new KN95 mask whose out-of-box filtration efficiency was measured to be $ 95 \pm 1\% $ (see Fig. \ref{fig:rechargen95recharge}(a)).  The mask was then washed at $ \sim 40^{\circ}$C in a conventional washing machine with detergent.  Such treatment would be expected to dissolve the lipid layer of the SARS-CoV-2 virus which causes COVID-19.  The mask was then air dried, and its efficiency was measured to be $ 75 \pm 1\% $ ( see Fig. \ref{fig:rechargen95recharge} (b) ).  The mask was then recharged for 60 minutes using the method of Figure \ref{fig:exptsetup}(b), following which its filtration efficiency was measured to be $ 95\% $ ( see Fig. \ref{fig:rechargen95recharge} (c) ). We then repeated this protocol, and found that the filtration efficiency reattained  $ 95 \pm 1\% $.  Figure \ref{fig:rechargen95recharge} (d) shows that the filtration efficiency of an exposed mask degrades only slightly, from $\sim 95 \pm 1\%$ to $\sim 92 \pm 1\%$, over the course of one day. This suggests that the use of sterilization procedures which do not cause structural damage to a mask coupled with our recharging protocol will produce a respirator which may be used multiple times with no sacrifice in filtration efficiency.

We have  verified that the recharging method works on a variety of N95 respirators, and that the filtration efficiency of degraded masks can be improved by charging, if not to brand-new efficiency.  This suggests that by using this method we should be able to determine the effect of various disinfection protocols on the structural integrity of different brands of mask.  In this context we note that a given sensitization method may affect different brands of masks very differently, as seen in Table \ref{tab:caption}.

\begin{figure}[t]
	\centering
		\includegraphics[width=1\linewidth]{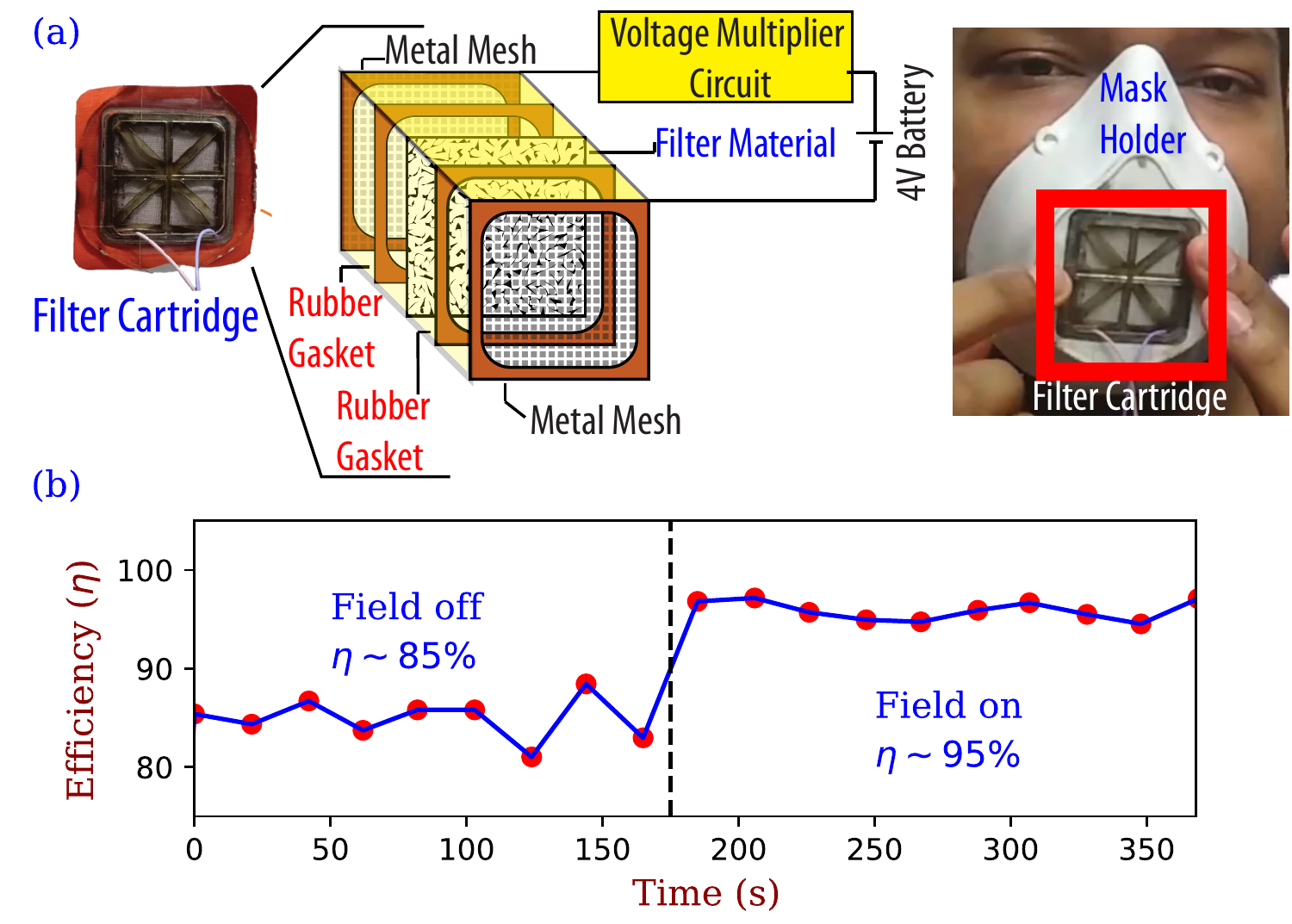}
	\caption{ (a) Schematic representation of the {in situ} continuously charged mask, whose cartridge  fits onto a 3D  printed housing.   (b)	 Upon applying a field, the efficiency of the cartridge improves from $85 \pm 1\% $to about $95 \pm 1 \%$. Filtration measured for 0. 3$\mu m$ sized particles.  }
	\label{fig:insitu}
\end{figure}

\subsection*{In-situ application of an electric field keeps the electret filter recharged}
\textcolor{color}{Today, N95 masks are being worn by health-care workers for extended periods of time. This gives rise to very humid conditions. Humidity is detrimental to electrostatics. Thus, during use, all electrostatics-based masks slowly lose their efficiency. A solution that can help replenish the lost charge on the masks in real time would be desirable. In this section we provide a proof-of-concept method of keeping the masks charged which comes  as a logical extension of our recharging method. } 

We tested a technique by which the filter material  maintains its charge, and thus its filtration efficiency. We do this by applying a high electric field in a current limited condition (very low current (few $\mu A$), so no risk of discharge or shocking) to the material {\it in-situ}.  Figure \ref{fig:insitu} shows a schematic of the {\it in-situ} setup: a layer of filtration material (polypropylene mesh)  cut from a standard N95 mask  including all its layers  serve as the filtration medium.  This filter is sandwiched between two porous metallic screens which are connected to a 4V battery whose voltage is multiplied to 2000V using a   standard voltage multiplier circuit. We use a rubber gasket on both sides of the mask material  to provide electrical insulation. The efficacy of the method is indicated in Figure \ref{fig:insitu}, where the filtration efficiency in the absence of an electric field is $ 85 \pm 1\% $, which rises to $ 95 \pm 1\% $ upon application of voltage.  We have verified that as long as the voltage is applied, the filtration efficiency remains high.

\textcolor{color} {Since the currents required are extremely small, a large battery is not required, and it is possible that a small compact and practical solution may be feasible.}

\subsection*{Conclusions}


Since the loss of electrical charge from the polypropylene filter layer in N95 masks is known to impact the filtration efficiency, we investigated the possibility of mask recharging for a few commercially available N95 masks using a simple laboratory setup. Our  results suggest that it is possible to recharge the masks post-sterilization and recover filtration efficiency. 
However, this is a promising development that merits further research as it may permit multiple or extended use in practical applications.  In particular, this method may allow for N95 masks to be used for a considerably longer period of time than is the current norm, which can have a significant effect in hospitals where mask supply is insufficient. Additionally, we envisage that our method may find applications in a variety of air filtration contexts.  We have focused in this paper on high efficiency respirators for use in preventing disease transmission, but we anticipate applications to HVAC and industrial filtration as well, where our recharging method would allow for the extended use of electrostatic filters, resulting in reduced cost and waste.  Furthermore, our in-situ field application makes possible high efficiency filtration with undiminished performance over time.  

\section*{Supplementary Material}

\subsection*{Construction of the mask particle filtration efficiency tester}

In this section we outline the design of the low-cost, compact, particle filtration efficiency test setup using a Plantower PMS 7003 particle concentration sensor air quality monitor chip and a ESP8266-based WiFi microcontroller that was used for the measurements of particle filtration efficiency reported in this work. All
the construction details, diagrams, source codes for the micro-controller and interface are available
open-source at the  \href{https://github.com/shescitech/TIFR_Mask_Efficiency} {GitHub repository}\footnote{https://github.com/shescitech/TIFR\_Mask\_Efficiency}.

A background of fine particles was generated using normal saline solution in a standard medical nebulizer to create a fine aerosol. Air is sucked through the particle counter using an oil-free diaphragm pump and the 
throughput of 0.3 $\mu m$ sized particles measured with and without a mask at the input. A
simple HTML-based web interface was used to read the particle counts and evaluate the filtration efficiency. We estimate a $ \pm 1\%$ accuracy for measurements made with this setup.

\subsubsection*{Design}

Any mask filtration efficiency test setup requires to subject a sheet of the mask material of fixed area
(or the entire mask, typically mounted on a human face mannequin), to a flux of 0.3 $\mu m$ sized
particles (for N95) and measure the throughput of such particles across the mask. Given the interest in air quality measurements, especially in polluted cities, air-quality- indicator
(AQI) systems are cheaply available and widely deployed. Most of these are designed to measure PM2.5 and PM10
levels (of 2.5 $\mu m$ and 10 $\mu m$ sized particles). However many such AQI systems are based on laser
particle counter sensors (like the Plantower PMS 1003/5003/7003 series \footnote{ \href{http://www.plantower.com/en/}{http://www.plantower.com/en/}}) that actually provide
measurements of particle counts at 0.3, 0.5, 1, 2.5, 5 and 10 $\mu m$ as standard outputs. These sensors
are optimized for the detection of 2.5 $\mu m$ sized particles and have been extensively tested \footnote{M.L. Zamora, F. Xiong, D. Gentner, B. Kerkez, J. Kohrman-Glaser, K. Koehler, Field and Laboratory Evaluations 	of the Low-Cost Plantower Particulate Matter Sensor, Environ. Sci. Technol., 53, 838–849 (2019)}

Though the sensitivity of the PMS 7003 at 0.3 $\mu m$ is considerably less than at 2.5 $\mu m$, it is reasonable enough
for reliable measurements. However, in any case the absolute sensitivity is not critical for a mask
test application as the parameter of interest is a ratio of measurements. The filtration efficiency,
$\eta_{mask}$, can be determined from the ratio of particles per unit time detected with, $ N_{mask} $, and without
the mask attached, $ N_{ambient} $, as 
\[
\eta_{mask} = 1-\left(\frac{N_{mask}}{N_{ambient}}\right)
\]
Our mask PFE test setup uses a Plantower PMS7003 sensor that is controlled via an ESP8266 WiFi
micro controller unit. We used a Wemos D1 R2 WiFi capable ESP8266 based development board for
the Arduino IDE. The setup is externally powered via a
standard micro USB port on the ESP8266 board. This is connected to the PMS7003 sensor using a 1.27mm
pitch IDC connector, which supplies power and is also used for data transfer via serial communication.

\begin{figure}
	\centering
	\includegraphics[width=1\linewidth]{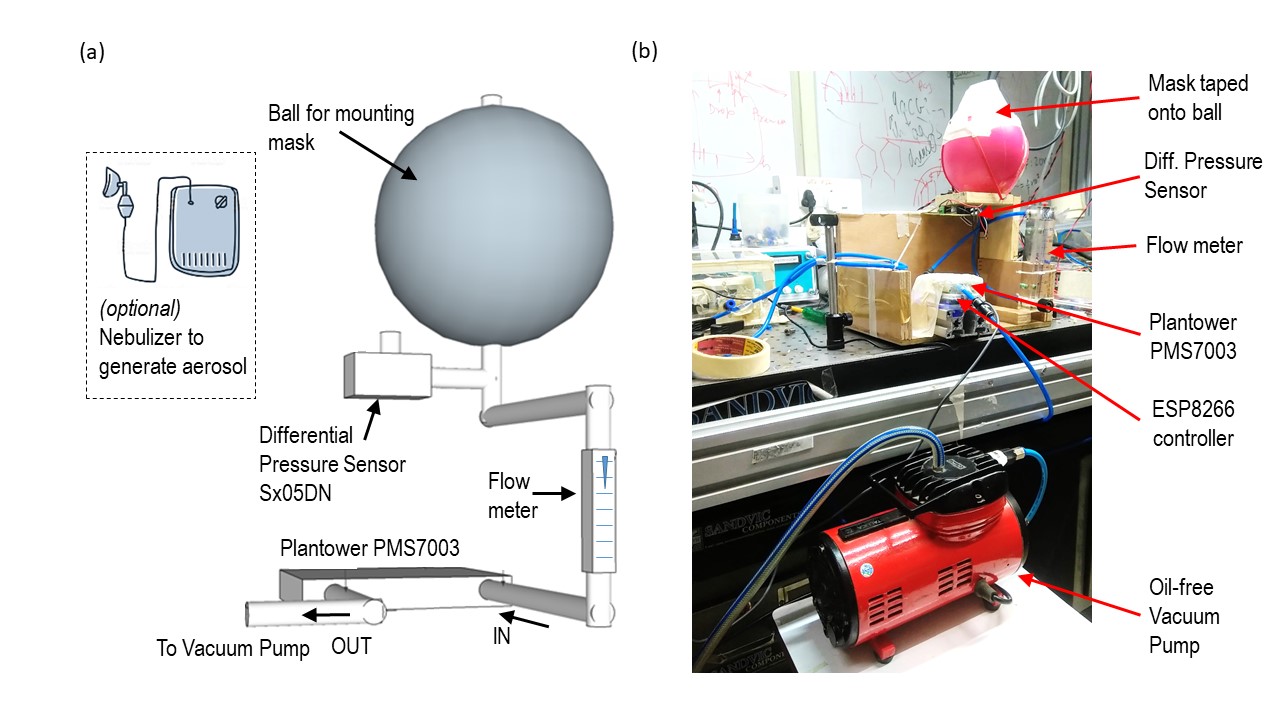}
	\includegraphics[width=1\linewidth]{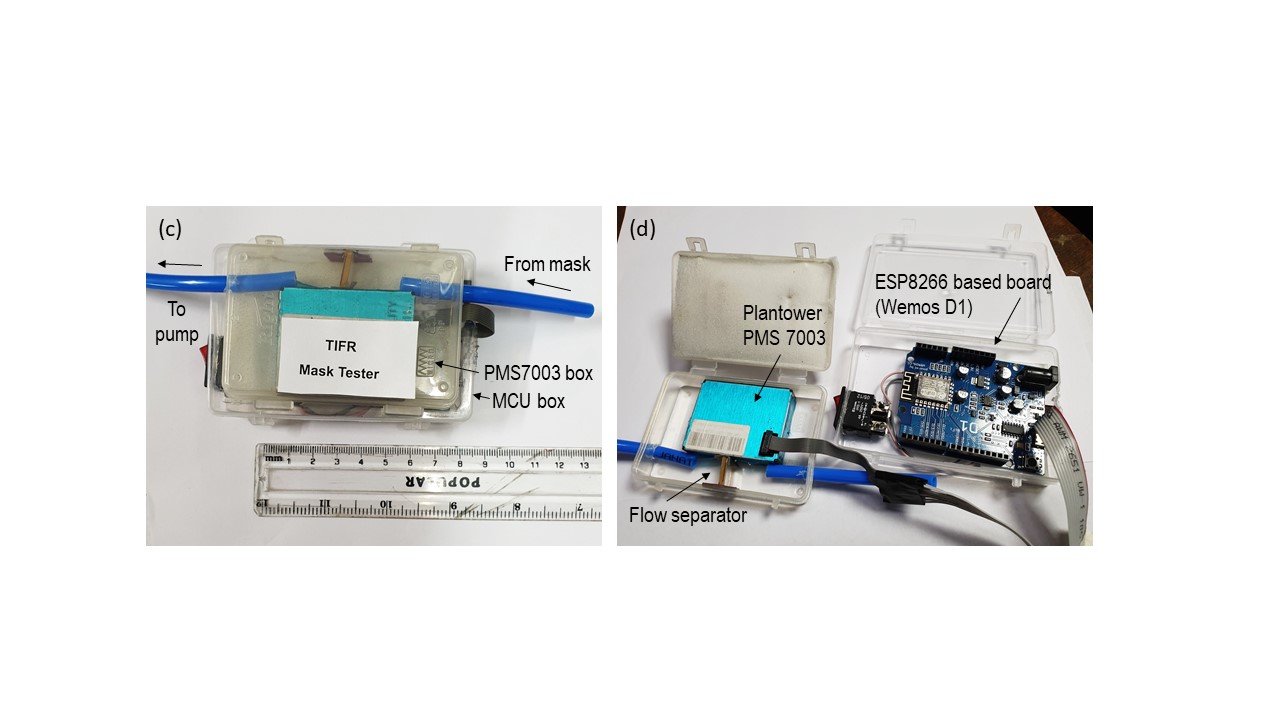}
	\caption{(a) Schematic diagram and (b) photograph of the PFE test setup, (c) view of the air quality sensor chip
		and control unit in small plastic boxes kept atop each other in a compact configuration (b) opened up view
		showing the PMS7003 chip and the ESP8266 based MCU board. The box edges and connector ports are sealed
		with tape/hot-glue to ensure that the setup is airtight.}
	\label{fig:sifigs1}
\end{figure}

A schematic of the setup and a photograph of the lab prototype are shown in Figs. \ref{fig:sifigs1} (a) and (b), respectively. The
PMS7003 is kept inside a small plastic box, with input and output provided via standard 6 mm plastic tubes.
We use 6 mm quick fit connectors all through the setup for ease of assembly. We use an oil-free
diaphragm pump (HSV-1, High Speed Appliances, Mumbai) that provides a maximum flow of 30 lpm.
The flow can be measured and controlled with a taper-tube flow meter. For most experiments we
used a flow of 10 lpm \footnote{We are aware of the ASTM test conditions for masks that specify a flow rate of 85 SLPM. However our setup
	is mainly aimed to be a tool for quick evaluation of mask PFE, e.g. to check if a claimed N95 mask really meets
	the specification, or to check before-and-after performance following decontamination treatments etc.}, similar to typical human breathing rates.

A critical design issue is that these sensors are designed to sample ambient air (with a small
internal fan) and do not have the ability to generate any suction pressure to draw air through a face
mask. Thus, some arrangement to force the air flow through the mask and past the sensor is needed, which was done with a small diaphragm pump. It is of utmost importance to ensure that the air being sucked through
the mask is appropriately directed into the input port of the particle counter sensor, to enable
reliable measurements to be made. Further, the vacuum pump has to sample air at the output port
of the sensor, and ensure that the air flow is such that it does not bypass the sensor. This can be
conveniently achieved by using an appropriately sized divider plate that sits tightly against the
PMS7003 sensor chip and the box. This serves as a barrier between output and input ports to
prevent direct mixing, and ensures that the air flows through the sensor. (See part marked as flow
separator in Fig. \ref{fig:sifigs1} (d)). It is also important that the PMS7003 sensor box and the input/output
connectors are completely sealed to prevent any external air being sucked into the box. We used a
combination of tape and hot-glue for this purpose.

\subsubsection*{Source of particles}
While typical room air has enough particles of 0.3$\mu m$ size to enable PFE measurements on N95
masks, air-conditioned laboratory environments often have filters that reduce the particulate count
significantly. In that case a much better signal to noise ratio in the measurements can be achieved
with intentionally enhanced particle counts in the local environment. This can easily be done using
normal saline (0.9\% NaCl solution) in a standard medical nebulizer. We can obtain a relatively constant background level with the nebulizer kept about 5 m away from the test setup.

\subsubsection*{Interfacing and user interface}

The codes required for programming the ESP8266 based MCU board to interface with the PMS7003
were written in the Arduino Integrated Development Environment (IDE) and allow the particle
count data from the PMS7003 to be streamed to a local http web server as well as to a local telnet server. A python code is used to
access the real time data from the telnet server and plot it, as well as apply the appropriate filters, analysis and PFE
calculation tools.

\begin{figure}
	\centering
	\includegraphics[width=0.7\linewidth]{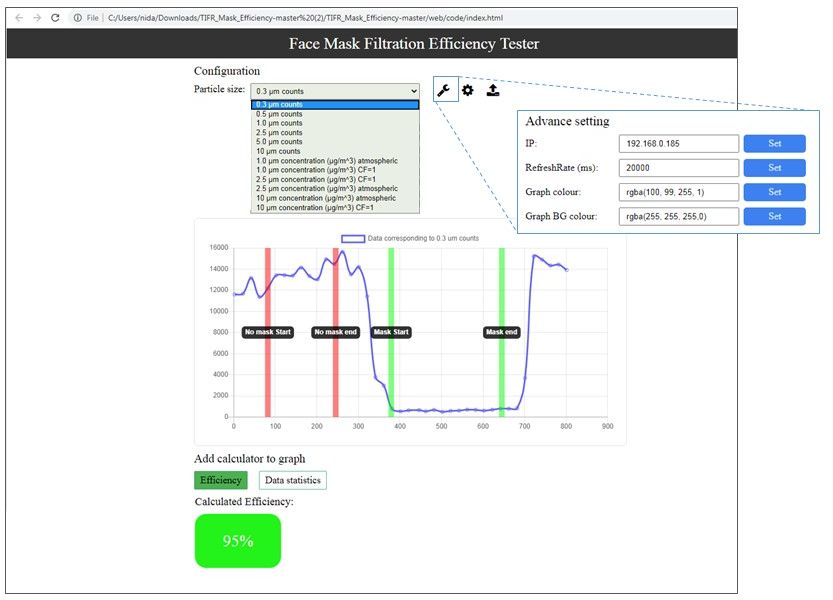}
	\caption{Screenshot of the face mask efficiency tester. A drop down menu allows different particle size channels
		to be chosen. Clicking on the wrench icon pops up an advanced settings window allowing other options to be
		changed. Clicking on the efficiency button provides 4 vertical bars that can be moved to demarcate regions of
		particle count without and with the mask, and the calculated PFE is displayed. Similarly the data statistics
		marker proves two vertical bars to demarcate a region for calculating the average and standard deviation of
		particle counts.}
	\label{fig:sifigs3}
\end{figure}

The data transmitted from the ESP8266 module to the web server is visualized and analyzed through a simple html interface that allows various parameters to be changed, including the particle size channel,
data acquisition rate, IP address etc. A snapshot of the mask tester user interface is shown in Fig. \ref{fig:sifigs3}.
To ease the calculation of the PFE an intuitive user interface with an in-built efficiency calculator has
also been provided.

Another useful feature of our code is the ability to modify parameters (e.g. change WiFi name and
password, acquisition rate in seconds etc.) via OTA (over the air) updates through the WiFi network
without having to connecting the MCU physically to a computer. Detailed procedures are available in
the relevant section of the GitHub repository.

The GitHub repository \href{https://github.com/shescitech/TIFR_Mask_Efficiency}{https://github.com/shescitech/TIFR\_Mask\_Efficiency}   has 4 sections:

\begin{enumerate}
	\item \textcolor{blue}{arduino} - instructions and code to integrate the Plantower PMS7003 with any ESP8266
	based board (Wemos D1 R2, Wemos D1 mini, Nodemcu etc) and stream the particle count
	data to a local http web server
	\item \textcolor{blue}{design} – overall project design guide and pictures
	\item \textcolor{blue}{python} – instructions and code to access to real time data from the particle counter
	\item \textcolor{blue} {web} – instructions and code to visualize the particles count and calculate mask PFE, data
	statistics, save data and graphs, upload and view previous measurements etc.
	
\end{enumerate}

\begin{acknowledgments}

 We thank Tata Memorial Hospital, Mumbai for sending us the samples of N95 / FFP2 masks for testing our setup.  We are grateful to Paul Chaikin for many useful and insightful discussions. EH  thanks C-CAMP Bengaluru   for financial support.  DL gratefully acknowledges the support of the US-Israel Binational Science Foundation (grant 2014713), the Israel Science Foundation (grant 1866/16), and a TIFR Visiting Professorship. We acknowledge support of the Department of Atomic Energy, Government of India, under Project No. 12-R\&D-TFR-5.10-0100. EH thanks  C-CAMP Bangalore, India for  financial support. 

\end{acknowledgments}

\medskip

\paragraph*{\textbf{Availability  of data : }} The data that supports the findings of this study are available within the article.   Additional  drawings and  interfacing codes of the mask tester  can be  found  in the \texttt{GitHub}   repository \cite{MAsk_tester}.

\bibliography{references}

\begin{thebibliography}{33}%
\makeatletter
\providecommand \@ifxundefined [1]{%
 \@ifx{#1\undefined}
}%
\providecommand \@ifnum [1]{%
 \ifnum #1\expandafter \@firstoftwo
 \else \expandafter \@secondoftwo
 \fi
}%
\providecommand \@ifx [1]{%
 \ifx #1\expandafter \@firstoftwo
 \else \expandafter \@secondoftwo
 \fi
}%
\providecommand \natexlab [1]{#1}%
\providecommand \enquote  [1]{``#1''}%
\providecommand \bibnamefont  [1]{#1}%
\providecommand \bibfnamefont [1]{#1}%
\providecommand \citenamefont [1]{#1}%
\providecommand \href@noop [0]{\@secondoftwo}%
\providecommand \href [0]{\begingroup \@sanitize@url \@href}%
\providecommand \@href[1]{\@@startlink{#1}\@@href}%
\providecommand \@@href[1]{\endgroup#1\@@endlink}%
\providecommand \@sanitize@url [0]{\catcode `\\12\catcode `\$12\catcode
  `\&12\catcode `\#12\catcode `\^12\catcode `\_12\catcode `\%12\relax}%
\providecommand \@@startlink[1]{}%
\providecommand \@@endlink[0]{}%
\providecommand \url  [0]{\begingroup\@sanitize@url \@url }%
\providecommand \@url [1]{\endgroup\@href {#1}{\urlprefix }}%
\providecommand \urlprefix  [0]{URL }%
\providecommand \Eprint [0]{\href }%
\providecommand \doibase [0]{http://dx.doi.org/}%
\providecommand \selectlanguage [0]{\@gobble}%
\providecommand \bibinfo  [0]{\@secondoftwo}%
\providecommand \bibfield  [0]{\@secondoftwo}%
\providecommand \translation [1]{[#1]}%
\providecommand \BibitemOpen [0]{}%
\providecommand \bibitemStop [0]{}%
\providecommand \bibitemNoStop [0]{.\EOS\space}%
\providecommand \EOS [0]{\spacefactor3000\relax}%
\providecommand \BibitemShut  [1]{\csname bibitem#1\endcsname}%
\let\auto@bib@innerbib\@empty
\bibitem [{\citenamefont {Dbouk}\ and\ \citenamefont
  {Drikakis}(2020{\natexlab{a}})}]{dbouk2020respiratory}%
  \BibitemOpen
  \bibfield  {author} {\bibinfo {author} {\bibfnamefont {T.}~\bibnamefont
  {Dbouk}}\ and\ \bibinfo {author} {\bibfnamefont {D.}~\bibnamefont
  {Drikakis}},\ }\bibfield  {title} {\enquote {\bibinfo {title} {On respiratory
  droplets and face masks},}\ }\href@noop {} {\bibfield  {journal} {\bibinfo
  {journal} {Physics of Fluids}\ }\textbf {\bibinfo {volume} {32}},\ \bibinfo
  {pages} {063303} (\bibinfo {year} {2020}{\natexlab{a}})}\BibitemShut
  {NoStop}%
\bibitem [{\citenamefont {Verma}, \citenamefont {Dhanak},\ and\ \citenamefont
  {Frankenfield}(2020)}]{verma2020visualizing}%
  \BibitemOpen
  \bibfield  {author} {\bibinfo {author} {\bibfnamefont {S.}~\bibnamefont
  {Verma}}, \bibinfo {author} {\bibfnamefont {M.}~\bibnamefont {Dhanak}}, \
  and\ \bibinfo {author} {\bibfnamefont {J.}~\bibnamefont {Frankenfield}},\
  }\bibfield  {title} {\enquote {\bibinfo {title} {Visualizing the
  effectiveness of face masks in obstructing respiratory jets},}\ }\href@noop
  {} {\bibfield  {journal} {\bibinfo  {journal} {Physics of Fluids}\ }\textbf
  {\bibinfo {volume} {32}},\ \bibinfo {pages} {061708} (\bibinfo {year}
  {2020})}\BibitemShut {NoStop}%
\bibitem [{\citenamefont {Busco}\ \emph {et~al.}(2020)\citenamefont {Busco},
  \citenamefont {Yang}, \citenamefont {Seo},\ and\ \citenamefont
  {Hassan}}]{busco2020sneezing}%
  \BibitemOpen
  \bibfield  {author} {\bibinfo {author} {\bibfnamefont {G.}~\bibnamefont
  {Busco}}, \bibinfo {author} {\bibfnamefont {S.~R.}\ \bibnamefont {Yang}},
  \bibinfo {author} {\bibfnamefont {J.}~\bibnamefont {Seo}}, \ and\ \bibinfo
  {author} {\bibfnamefont {Y.~A.}\ \bibnamefont {Hassan}},\ }\bibfield  {title}
  {\enquote {\bibinfo {title} {Sneezing and asymptomatic virus transmission},}\
  }\href@noop {} {\bibfield  {journal} {\bibinfo  {journal} {Physics of
  Fluids}\ }\textbf {\bibinfo {volume} {32}},\ \bibinfo {pages} {073309}
  (\bibinfo {year} {2020})}\BibitemShut {NoStop}%
\bibitem [{\citenamefont {Brosseau}\ and\ \citenamefont
  {Ann}(2009)}]{brosseau2009n95}%
  \BibitemOpen
  \bibfield  {author} {\bibinfo {author} {\bibfnamefont {L.}~\bibnamefont
  {Brosseau}}\ and\ \bibinfo {author} {\bibfnamefont {R.~B.}\ \bibnamefont
  {Ann}},\ }\bibfield  {title} {\enquote {\bibinfo {title} {N95 respirators and
  surgical masks},}\ }\href@noop {} {\bibfield  {journal} {\bibinfo  {journal}
  {Centers for Disease Control and Prevention}\ } (\bibinfo {year}
  {2009})}\BibitemShut {NoStop}%
\bibitem [{\citenamefont {for Disease~Control}, \citenamefont {Prevention}\
  \emph {et~al.}(2020)\citenamefont {for Disease~Control}, \citenamefont
  {Prevention} \emph {et~al.}}]{centers2020decontamination}%
  \BibitemOpen
  \bibfield  {author} {\bibinfo {author} {\bibfnamefont {C.}~\bibnamefont {for
  Disease~Control}}, \bibinfo {author} {\bibnamefont {Prevention}},  \emph
  {et~al.},\ }\bibfield  {title} {\enquote {\bibinfo {title} {Decontamination
  and reuse of filtering facepiece respirators},}\ }\href@noop {} {\bibfield
  {journal} {\bibinfo  {journal} {Reviewed April}\ }\textbf {\bibinfo {volume}
  {9}} (\bibinfo {year} {2020})}\BibitemShut {NoStop}%
\bibitem [{\citenamefont {Dbouk}\ and\ \citenamefont
  {Drikakis}(2020{\natexlab{b}})}]{dbouk2020coughing}%
  \BibitemOpen
  \bibfield  {author} {\bibinfo {author} {\bibfnamefont {T.}~\bibnamefont
  {Dbouk}}\ and\ \bibinfo {author} {\bibfnamefont {D.}~\bibnamefont
  {Drikakis}},\ }\bibfield  {title} {\enquote {\bibinfo {title} {On coughing
  and airborne droplet transmission to humans},}\ }\href@noop {} {\bibfield
  {journal} {\bibinfo  {journal} {Physics of Fluids}\ }\textbf {\bibinfo
  {volume} {32}},\ \bibinfo {pages} {053310} (\bibinfo {year}
  {2020}{\natexlab{b}})}\BibitemShut {NoStop}%
\bibitem [{\citenamefont {Fischer}\ \emph {et~al.}(2020)\citenamefont {Fischer}
  \emph {et~al.}}]{fischer2020assessment}%
  \BibitemOpen
  \bibfield  {author} {\bibinfo {author} {\bibfnamefont {R.}~\bibnamefont
  {Fischer}} \emph {et~al.},\ }\bibfield  {title} {\enquote {\bibinfo {title}
  {Assessment of n95 respirator decontamination and re-use for sars-cov-2},}\
  }\href@noop {} {\bibfield  {journal} {\bibinfo  {journal} {medRxiv}\ }
  (\bibinfo {year} {2020})}\BibitemShut {NoStop}%
\bibitem [{\citenamefont {Liao}\ \emph {et~al.}(2020)\citenamefont {Liao} \emph
  {et~al.}}]{liao2020can}%
  \BibitemOpen
  \bibfield  {author} {\bibinfo {author} {\bibfnamefont {L.}~\bibnamefont
  {Liao}} \emph {et~al.},\ }\bibfield  {title} {\enquote {\bibinfo {title} {Can
  n95 respirators be reused after disinfection? how many times?}}\ }\href
  {\doibase 10.1021/acsnano.0c03597} {\bibfield  {journal} {\bibinfo  {journal}
  {ACS Nano}\ }\textbf {\bibinfo {volume} {14}},\ \bibinfo {pages} {6348--6356}
  (\bibinfo {year} {2020})},\ \bibinfo {note} {pMID: 32368894},\ \Eprint
  {http://arxiv.org/abs/https://doi.org/10.1021/acsnano.0c03597}
  {https://doi.org/10.1021/acsnano.0c03597} \BibitemShut {NoStop}%
\bibitem [{\citenamefont {Kumar}\ \emph {et~al.}(2020)\citenamefont {Kumar}
  \emph {et~al.}}]{kumar2020n95}%
  \BibitemOpen
  \bibfield  {author} {\bibinfo {author} {\bibfnamefont {A.}~\bibnamefont
  {Kumar}} \emph {et~al.},\ }\bibfield  {title} {\enquote {\bibinfo {title}
  {N95 mask decontamination using standard hospital sterilization
  technologies},}\ }\href {\doibase 10.1101/2020.04.05.20049346} {\bibfield
  {journal} {\bibinfo  {journal} {medRxiv}\ } (\bibinfo {year} {2020}),\
  10.1101/2020.04.05.20049346},\ \Eprint
  {http://arxiv.org/abs/https://www.medrxiv.org/content/early/2020/04/20/2020.04.05.20049346.full.pdf}
  {https://www.medrxiv.org/content/early/2020/04/20/2020.04.05.20049346.full.pdf}
  \BibitemShut {NoStop}%
\bibitem [{\citenamefont {O'Hearn}\ \emph
  {et~al.}(2020{\natexlab{a}})\citenamefont {O'Hearn} \emph
  {et~al.}}]{2020decontaminating}%
  \BibitemOpen
  \bibfield  {author} {\bibinfo {author} {\bibfnamefont {K.}~\bibnamefont
  {O'Hearn}} \emph {et~al.},\ }\bibfield  {title} {\enquote {\bibinfo {title}
  {Decontaminating n95 and sn95 masks with ultraviolet germicidal irradiation
  does not impair mask efficacy and safety},}\ }\href {\doibase
  https://doi.org/10.1016/j.jhin.2020.07.014} {\bibfield  {journal} {\bibinfo
  {journal} {Journal of Hospital Infection}\ }\textbf {\bibinfo {volume}
  {106}},\ \bibinfo {pages} {163} (\bibinfo {year}
  {2020}{\natexlab{a}})}\BibitemShut {NoStop}%
\bibitem [{\citenamefont {Li}\ \emph {et~al.}(2020)\citenamefont {Li},
  \citenamefont {Cadnum}, \citenamefont {Redmond}, \citenamefont {Jones},\ and\
  \citenamefont {Donskey}}]{li2020s}%
  \BibitemOpen
  \bibfield  {author} {\bibinfo {author} {\bibfnamefont {D.~F.}\ \bibnamefont
  {Li}}, \bibinfo {author} {\bibfnamefont {J.~L.}\ \bibnamefont {Cadnum}},
  \bibinfo {author} {\bibfnamefont {S.~N.}\ \bibnamefont {Redmond}}, \bibinfo
  {author} {\bibfnamefont {L.~D.}\ \bibnamefont {Jones}}, \ and\ \bibinfo
  {author} {\bibfnamefont {C.~J.}\ \bibnamefont {Donskey}},\ }\bibfield
  {title} {\enquote {\bibinfo {title} {It's not the heat, it's the humidity:
  Effectiveness of a rice cooker-steamer for decontamination of cloth and
  surgical face masks and n95 respirators},}\ }\href {\doibase
  https://doi.org/10.1016/j.ajic.2020.04.012} {\bibfield  {journal} {\bibinfo
  {journal} {American Journal of Infection Control}\ }\textbf {\bibinfo
  {volume} {48}},\ \bibinfo {pages} {854 -- 855} (\bibinfo {year}
  {2020})}\BibitemShut {NoStop}%
\bibitem [{\citenamefont {Mackenzie}(2020)}]{mackenzie2020reuse}%
  \BibitemOpen
  \bibfield  {author} {\bibinfo {author} {\bibfnamefont {D.}~\bibnamefont
  {Mackenzie}},\ }\bibfield  {title} {\enquote {\bibinfo {title} {Reuse of n95
  masks},}\ }\href {\doibase https://doi.org/10.1016/j.eng.2020.04.003}
  {\bibfield  {journal} {\bibinfo  {journal} {Engineering}\ }\textbf {\bibinfo
  {volume} {6}},\ \bibinfo {pages} {593 -- 596} (\bibinfo {year}
  {2020})}\BibitemShut {NoStop}%
\bibitem [{\citenamefont {Ma}\ \emph {et~al.}()\citenamefont {Ma} \emph
  {et~al.}}]{ma2020decontamination}%
  \BibitemOpen
  \bibfield  {author} {\bibinfo {author} {\bibfnamefont {Q.-X.}\ \bibnamefont
  {Ma}} \emph {et~al.},\ }\bibfield  {title} {\enquote {\bibinfo {title}
  {Decontamination of face masks with steam for mask reuse in fighting the
  pandemic covid-19: Experimental supports},}\ }\href {\doibase
  10.1002/jmv.25921} {\bibfield  {journal} {\bibinfo  {journal} {Journal of
  Medical Virology}\ }\textbf {\bibinfo {volume} {n/a}},\ 10.1002/jmv.25921},\
  \Eprint
  {http://arxiv.org/abs/https://onlinelibrary.wiley.com/doi/pdf/10.1002/jmv.25921}
  {https://onlinelibrary.wiley.com/doi/pdf/10.1002/jmv.25921} \BibitemShut
  {NoStop}%
\bibitem [{\citenamefont {O'Hearn}\ \emph
  {et~al.}(2020{\natexlab{b}})\citenamefont {O'Hearn} \emph
  {et~al.}}]{o2020efficacy}%
  \BibitemOpen
  \bibfield  {author} {\bibinfo {author} {\bibnamefont {O'Hearn}} \emph
  {et~al.},\ }\bibfield  {title} {\enquote {\bibinfo {title} {Efficacy and
  safety of disinfectants for decontamination of n95 and sn95 filtering
  facepiece respirators: A systematic review},}\ }\href@noop {} {\  (\bibinfo
  {year} {2020}{\natexlab{b}})}\BibitemShut {NoStop}%
\bibitem [{MAs()}]{MAsk_tester}%
  \BibitemOpen
  \href@noop {} {\enquote {\bibinfo {title} {Tifr mask tester \texttt{GitHub}
  repository},}\ }\bibinfo {howpublished}
  {\url{https://github.com/shescitech/TIFR_Mask_Efficiency}}\BibitemShut
  {NoStop}%
\bibitem [{\citenamefont {Thakur}, \citenamefont {Das},\ and\ \citenamefont
  {Das}(2013)}]{thakur2013electret}%
  \BibitemOpen
  \bibfield  {author} {\bibinfo {author} {\bibfnamefont {R.}~\bibnamefont
  {Thakur}}, \bibinfo {author} {\bibfnamefont {D.}~\bibnamefont {Das}}, \ and\
  \bibinfo {author} {\bibfnamefont {A.}~\bibnamefont {Das}},\ }\bibfield
  {title} {\enquote {\bibinfo {title} {Electret air filters},}\ }\href@noop {}
  {\bibfield  {journal} {\bibinfo  {journal} {Separation \& Purification
  Reviews}\ }\textbf {\bibinfo {volume} {42}},\ \bibinfo {pages} {87--129}
  (\bibinfo {year} {2013})}\BibitemShut {NoStop}%
\bibitem [{\citenamefont {Kumar}, \citenamefont {Bhattacharya},\ and\
  \citenamefont {Ghosh}(2013)}]{kumar2013weak}%
  \BibitemOpen
  \bibfield  {author} {\bibinfo {author} {\bibfnamefont {D.}~\bibnamefont
  {Kumar}}, \bibinfo {author} {\bibfnamefont {S.}~\bibnamefont {Bhattacharya}},
  \ and\ \bibinfo {author} {\bibfnamefont {S.}~\bibnamefont {Ghosh}},\
  }\bibfield  {title} {\enquote {\bibinfo {title} {Weak adhesion at the
  mesoscale: particles at an interface},}\ }\href@noop {} {\bibfield  {journal}
  {\bibinfo  {journal} {Soft Matter}\ }\textbf {\bibinfo {volume} {9}},\
  \bibinfo {pages} {6618--6633} (\bibinfo {year} {2013})}\BibitemShut {NoStop}%
\bibitem [{\citenamefont {Finlay}(2001)}]{finlay2001mechanics}%
  \BibitemOpen
  \bibfield  {author} {\bibinfo {author} {\bibfnamefont {W.~H.}\ \bibnamefont
  {Finlay}},\ }\href@noop {} {\emph {\bibinfo {title} {The mechanics of inhaled
  pharmaceutical aerosols: an introduction}}}\ (\bibinfo  {publisher} {Academic
  press},\ \bibinfo {year} {2001})\BibitemShut {NoStop}%
\bibitem [{\citenamefont {Lee}\ and\ \citenamefont
  {Liu}(1980)}]{lee1980minimum}%
  \BibitemOpen
  \bibfield  {author} {\bibinfo {author} {\bibfnamefont {K.}~\bibnamefont
  {Lee}}\ and\ \bibinfo {author} {\bibfnamefont {B.}~\bibnamefont {Liu}},\
  }\bibfield  {title} {\enquote {\bibinfo {title} {On the minimum efficiency
  and the most penetrating particle size for fibrous filters},}\ }\href@noop {}
  {\bibfield  {journal} {\bibinfo  {journal} {Journal of the Air Pollution
  Control Association}\ }\textbf {\bibinfo {volume} {30}},\ \bibinfo {pages}
  {377--381} (\bibinfo {year} {1980})}\BibitemShut {NoStop}%
\bibitem [{\citenamefont {Frederick}(1974)}]{frederick1974some}%
  \BibitemOpen
  \bibfield  {author} {\bibinfo {author} {\bibfnamefont {E.~R.}\ \bibnamefont
  {Frederick}},\ }\bibfield  {title} {\enquote {\bibinfo {title} {Some effects
  of electrostatic charges in fabric filtration},}\ }\href@noop {} {\bibfield
  {journal} {\bibinfo  {journal} {Journal of the Air Pollution Control
  Association}\ }\textbf {\bibinfo {volume} {24}},\ \bibinfo {pages}
  {1164--1168} (\bibinfo {year} {1974})}\BibitemShut {NoStop}%
\bibitem [{\citenamefont {Sessler}(1980)}]{sessler1980physical}%
  \BibitemOpen
  \bibfield  {author} {\bibinfo {author} {\bibfnamefont {G.~M.}\ \bibnamefont
  {Sessler}},\ }\bibfield  {title} {\enquote {\bibinfo {title} {Physical
  principles of electrets},}\ }in\ \href@noop {} {\emph {\bibinfo {booktitle}
  {Electrets}}}\ (\bibinfo  {publisher} {Springer},\ \bibinfo {year} {1980})\
  pp.\ \bibinfo {pages} {13--80}\BibitemShut {NoStop}%
\bibitem [{\citenamefont {Zhang}\ \emph {et~al.}(2018)\citenamefont {Zhang},
  \citenamefont {Liu}, \citenamefont {Zhang}, \citenamefont {Huang},\ and\
  \citenamefont {Jin}}]{zhang2018design}%
  \BibitemOpen
  \bibfield  {author} {\bibinfo {author} {\bibfnamefont {H.}~\bibnamefont
  {Zhang}}, \bibinfo {author} {\bibfnamefont {J.}~\bibnamefont {Liu}}, \bibinfo
  {author} {\bibfnamefont {X.}~\bibnamefont {Zhang}}, \bibinfo {author}
  {\bibfnamefont {C.}~\bibnamefont {Huang}}, \ and\ \bibinfo {author}
  {\bibfnamefont {X.}~\bibnamefont {Jin}},\ }\bibfield  {title} {\enquote
  {\bibinfo {title} {Design of electret polypropylene melt blown air filtration
  material containing nucleating agent for effective pm2. 5 capture},}\
  }\href@noop {} {\bibfield  {journal} {\bibinfo  {journal} {RSC Advances}\
  }\textbf {\bibinfo {volume} {8}},\ \bibinfo {pages} {7932--7941} (\bibinfo
  {year} {2018})}\BibitemShut {NoStop}%
\bibitem [{\citenamefont {Kilic}, \citenamefont {Shim},\ and\ \citenamefont
  {Pourdeyhimi}(2015)}]{kilic2015electrostatic}%
  \BibitemOpen
  \bibfield  {author} {\bibinfo {author} {\bibfnamefont {A.}~\bibnamefont
  {Kilic}}, \bibinfo {author} {\bibfnamefont {E.}~\bibnamefont {Shim}}, \ and\
  \bibinfo {author} {\bibfnamefont {B.}~\bibnamefont {Pourdeyhimi}},\
  }\bibfield  {title} {\enquote {\bibinfo {title} {Electrostatic capture
  efficiency enhancement of polypropylene electret filters with barium
  titanate},}\ }\href {\doibase 10.1080/02786826.2015.1061649} {\bibfield
  {journal} {\bibinfo  {journal} {Aerosol Science and Technology}\ }\textbf
  {\bibinfo {volume} {49}},\ \bibinfo {pages} {666--673} (\bibinfo {year}
  {2015})}\BibitemShut {NoStop}%
\bibitem [{\citenamefont {Pai}\ and\ \citenamefont
  {Springett}(1993)}]{pai1993physics}%
  \BibitemOpen
  \bibfield  {author} {\bibinfo {author} {\bibfnamefont {D.~M.}\ \bibnamefont
  {Pai}}\ and\ \bibinfo {author} {\bibfnamefont {B.~E.}\ \bibnamefont
  {Springett}},\ }\bibfield  {title} {\enquote {\bibinfo {title} {Physics of
  electrophotography},}\ }\href@noop {} {\bibfield  {journal} {\bibinfo
  {journal} {Reviews of Modern Physics}\ }\textbf {\bibinfo {volume} {65}},\
  \bibinfo {pages} {163} (\bibinfo {year} {1993})}\BibitemShut {NoStop}%
\bibitem [{\citenamefont {Gross}\ and\ \citenamefont
  {De~Moraes}(1962)}]{gross1962gamma}%
  \BibitemOpen
  \bibfield  {author} {\bibinfo {author} {\bibfnamefont {B.}~\bibnamefont
  {Gross}}\ and\ \bibinfo {author} {\bibfnamefont {R.}~\bibnamefont
  {De~Moraes}},\ }\bibfield  {title} {\enquote {\bibinfo {title} {Gamma
  irradiation effects on electrets},}\ }\href@noop {} {\bibfield  {journal}
  {\bibinfo  {journal} {Physical Review}\ }\textbf {\bibinfo {volume} {126}},\
  \bibinfo {pages} {930} (\bibinfo {year} {1962})}\BibitemShut {NoStop}%
\bibitem [{\citenamefont {Gross}(1958)}]{gross1958irradiation}%
  \BibitemOpen
  \bibfield  {author} {\bibinfo {author} {\bibfnamefont {B.}~\bibnamefont
  {Gross}},\ }\bibfield  {title} {\enquote {\bibinfo {title} {Irradiation
  effects in plexiglas},}\ }\href@noop {} {\bibfield  {journal} {\bibinfo
  {journal} {Journal of Polymer Science}\ }\textbf {\bibinfo {volume} {27}},\
  \bibinfo {pages} {135--143} (\bibinfo {year} {1958})}\BibitemShut {NoStop}%
\bibitem [{\citenamefont {McCarty}, \citenamefont {Winkleman},\ and\
  \citenamefont {Whitesides}(2007)}]{mccarty2007ionic}%
  \BibitemOpen
  \bibfield  {author} {\bibinfo {author} {\bibfnamefont {L.~S.}\ \bibnamefont
  {McCarty}}, \bibinfo {author} {\bibfnamefont {A.}~\bibnamefont {Winkleman}},
  \ and\ \bibinfo {author} {\bibfnamefont {G.~M.}\ \bibnamefont {Whitesides}},\
  }\bibfield  {title} {\enquote {\bibinfo {title} {Ionic electrets:
  electrostatic charging of surfaces by transferring mobile ions upon
  contact},}\ }\href@noop {} {\bibfield  {journal} {\bibinfo  {journal}
  {Journal of the American Chemical Society}\ }\textbf {\bibinfo {volume}
  {129}},\ \bibinfo {pages} {4075--4088} (\bibinfo {year} {2007})}\BibitemShut
  {NoStop}%
\bibitem [{\citenamefont {Zhao}\ \emph {et~al.}(2020)\citenamefont {Zhao} \emph
  {et~al.}}]{zhao2020household}%
  \BibitemOpen
  \bibfield  {author} {\bibinfo {author} {\bibfnamefont {M.}~\bibnamefont
  {Zhao}} \emph {et~al.},\ }\bibfield  {title} {\enquote {\bibinfo {title}
  {Household materials selection for homemade cloth face coverings and their
  filtration efficiency enhancement with triboelectric charging},}\ }\href
  {\doibase 10.1021/acs.nanolett.0c02211} {\bibfield  {journal} {\bibinfo
  {journal} {Nano Letters}\ }\textbf {\bibinfo {volume} {20}},\ \bibinfo
  {pages} {5544} (\bibinfo {year} {2020})}\BibitemShut {NoStop}%
\bibitem [{\citenamefont {Konda}\ \emph {et~al.}(2020)\citenamefont {Konda}
  \emph {et~al.}}]{konda2020aerosol}%
  \BibitemOpen
  \bibfield  {author} {\bibinfo {author} {\bibfnamefont {A.}~\bibnamefont
  {Konda}} \emph {et~al.},\ }\bibfield  {title} {\enquote {\bibinfo {title}
  {Aerosol filtration efficiency of common fabrics used in respiratory cloth
  masks},}\ }\href@noop {} {\bibfield  {journal} {\bibinfo  {journal} {ACS
  nano}\ }\textbf {\bibinfo {volume} {14}},\ \bibinfo {pages} {6339--6347}
  (\bibinfo {year} {2020})}\BibitemShut {NoStop}%
\bibitem [{\citenamefont {Chudleigh}, \citenamefont {Collins},\ and\
  \citenamefont {Hancock}(1973)}]{chudleigh1973stability}%
  \BibitemOpen
  \bibfield  {author} {\bibinfo {author} {\bibfnamefont {P.}~\bibnamefont
  {Chudleigh}}, \bibinfo {author} {\bibfnamefont {R.}~\bibnamefont {Collins}},
  \ and\ \bibinfo {author} {\bibfnamefont {G.}~\bibnamefont {Hancock}},\
  }\bibfield  {title} {\enquote {\bibinfo {title} {Stability of liquid charged
  electrets},}\ }\href@noop {} {\bibfield  {journal} {\bibinfo  {journal}
  {Applied physics letters}\ }\textbf {\bibinfo {volume} {23}},\ \bibinfo
  {pages} {211--212} (\bibinfo {year} {1973})}\BibitemShut {NoStop}%
\bibitem [{\citenamefont {Ferron}\ \emph {et~al.}(1997)\citenamefont {Ferron},
  \citenamefont {Roth}, \citenamefont {Busch},\ and\ \citenamefont
  {Karg}}]{ferron1997estimation}%
  \BibitemOpen
  \bibfield  {author} {\bibinfo {author} {\bibfnamefont {G.}~\bibnamefont
  {Ferron}}, \bibinfo {author} {\bibfnamefont {C.}~\bibnamefont {Roth}},
  \bibinfo {author} {\bibfnamefont {B.}~\bibnamefont {Busch}}, \ and\ \bibinfo
  {author} {\bibfnamefont {E.}~\bibnamefont {Karg}},\ }\bibfield  {title}
  {\enquote {\bibinfo {title} {Estimation of the size distribution of aerosols
  produced by jet nebulizers as a function of time},}\ }\href@noop {}
  {\bibfield  {journal} {\bibinfo  {journal} {Journal of aerosol science}\
  }\textbf {\bibinfo {volume} {28}},\ \bibinfo {pages} {805--819} (\bibinfo
  {year} {1997})}\BibitemShut {NoStop}%
\bibitem [{\citenamefont {Foss}\ and\ \citenamefont
  {Dannhauser}(1963)}]{foss1963electrical}%
  \BibitemOpen
  \bibfield  {author} {\bibinfo {author} {\bibfnamefont {R.~A.}\ \bibnamefont
  {Foss}}\ and\ \bibinfo {author} {\bibfnamefont {W.}~\bibnamefont
  {Dannhauser}},\ }\bibfield  {title} {\enquote {\bibinfo {title} {Electrical
  conductivity of polypropylene},}\ }\href@noop {} {\bibfield  {journal}
  {\bibinfo  {journal} {Journal of Applied Polymer Science}\ }\textbf {\bibinfo
  {volume} {7}},\ \bibinfo {pages} {1015--1022} (\bibinfo {year}
  {1963})}\BibitemShut {NoStop}%
\bibitem [{\citenamefont {Okoniewski}\ and\ \citenamefont
  {Perlman}(1994)}]{okoniewski1994hopping}%
  \BibitemOpen
  \bibfield  {author} {\bibinfo {author} {\bibfnamefont {A.}~\bibnamefont
  {Okoniewski}}\ and\ \bibinfo {author} {\bibfnamefont {M.}~\bibnamefont
  {Perlman}},\ }\bibfield  {title} {\enquote {\bibinfo {title} {Hopping
  conduction in “pure” polypropylene},}\ }\href@noop {} {\bibfield
  {journal} {\bibinfo  {journal} {Journal of Polymer Science Part B: Polymer
  Physics}\ }\textbf {\bibinfo {volume} {32}},\ \bibinfo {pages} {2413--2420}
  (\bibinfo {year} {1994})}\BibitemShut {NoStop}%
\end{thebibliography}%

\end{document}